\documentclass[12pt]{iopart}
\usepackage[utf8]{inputenc}
\usepackage{cite}
\usepackage{graphicx}
\usepackage{url}
\usepackage[T1]{fontenc}

\begin{document}

\title[Ciensación: sentir la ciencia]{Ciensación: sentir la ciencia}

\author{Marcos Henrique Abreu de Oliveira$^1$ and Robert Fischer$^2$}
\address{$^1$Instituto Federal de Alagoas, Maceió, AL, Brazil\\$^2$Universidade Tecnológica Federal do Paraná, Curitiba, PR, Brazil}
\ead{r@robertfischer.eu}

\begin{abstract}
Ciensación es un compilado abierto y en línea de experimentos prácticos que ha sido creado para convencer a los docentes latinoamericanos de que la mejor forma de vivir la ciencia es en primera persona. Permitir que los alumnos realicen experimentos de forma autónoma y en pequeños grupos es una tarea desafiante para los educadores de estos países. Analizamos los motivos por los cuales los docentes se resisten a utilizar experimentos prácticos en clase y debatimos cómo se implementó Ciensación para sortear esos obstáculos. El resultado fue la creación de actividades de investigación para los alumnos, que pueden integrarse fácilmente a la educación científica formal de las escuelas, con el objetivo de desarrollar habilidades científicas y, más importante aún, para que los jóvenes puedan ver la ciencia como una actividad creativa e interesante. \textit{ [Traducción del original inglés \guillemotleft Ciênsação, gaining a feeling for sciences\guillemotright ~por Catalina Estrada]}
\end{abstract}


\section{Experimentos prácticos en escuelas brasileñas}
Cuando Richard P. Feynman, un apasionado profesor de física, visitó Brasil en 1951, observó con gran sorpresa \guillemotleft niños de escuela primaria comprando libros de física en las librerías\guillemotright ~\cite{feynman}. Sin embargo, después de dar clase un semestre en una universidad en Rio de Janeiro, llegó a la conclusión de que \guillemotleft en Brasil no se enseña ciencia\guillemotright . En sus memorias escribió: \guillemotleft Luego de mucha investigación, finalmente descubrí que los alumnos habían memorizado todo, pero no sabían qué significaba todo eso\guillemotright . Aunque podían \guillemotleft repetir palabra por palabra\guillemotright ~conceptos de física avanzada, no podían relacionarlos a problemas del mundo real, ni aplicarlos en experimentos sencillos. Feynman abordó este tema en una charla que dio al final del semestre: \guillemotleft Por último, dije que no podía entender cómo alguien puede ser educado en ese sistema autopropagante, en el que la gente aprueba exámenes y enseñan a los demás a aprobar exámenes, pero en el que nadie sabe nada\guillemotright ~\cite{feynman}.

En los 65 años que pasaron desde la visita de Feynman, el sistema educativo brasileño ha mejorado significativamente. Hoy en día, las instituciones educativas hacen más hincapié en la comprensión y en la evidencia experimental. Las iniciativas para instalar el debate de los fenómenos naturales en un contexto relevante y significativo comenzaron a generar cambios en el currículo y en los libros didácticos y están llegando a las salas de aula. Sin embargo, los experimentos prácticos tal como los propone Feynman, aún son poco vistos en escuelas brasileñas. Este tipo de actividades centradas en el alumno, en las que en pequeños grupos se exploran fenómenos naturales con las propias manos y la propia mente, no se utilizan simplemente \guillemotleft por diversión\guillemotright ~o para despertar el interés de los jóvenes por la ciencia, sino que son un aporte esencial para la evolución de los estudiantes estimulando el desarrollo de habilidades y la autonomía intelectual basada en la evidencia. Para comprender por qué los docentes en Brasil aún no se deciden a implementar este método didáctico comprobado~\cite{Hattie,stohr}, es importante observar la situación local y las dificultades que los profesores enfrentan al realizar experimentos en clase. Durante varios años, el proyecto Ciensación ha estudiado este entorno de aprendizaje con relación a las políticas educativas, la formación docente y las prácticas cotidianas de enseñanza en escuelas primarias y secundarias, para luego desarrollar un método para abordar los obstáculos comunes y convencer a los docentes de realizar experimentos prácticos en clase. 

Ya en 1997, el Ministerio de Educación publicó en el currículo escolar nacional~\cite{PCN} la siguiente recomendación: \guillemotleft Un experimento se torna más interesante para los alumnos cuando ellos participan en su planificación, lo llevan a cabo ellos mismos, manipulan los aparatos y luego debaten sobre los resultados\guillemotright . Fueron pocos los docentes que siguieron este consejo en aquella época. En los últimos años, sin embargo, las políticas educativas resaltan cada vez más la importancia de estimular el desarrollo de habilidades, tales como pensamiento crítico y analítico, trabajo en equipo y buena comunicación~\cite{bruns}, que están muy relacionadas con los experimentos prácticos~\cite{kyle,stohr}. Un buen ejemplo es el ENEM, un examen de admisión para universidades públicas de todo el país, que ha sido reformulado para evaluar habilidades y competencias. La memorización de conocimientos fácticos aún desempeña un papel importante, pero responder preguntas científicas ahora exige una verdadera comprensión. En consecuencia, los docentes tienen que adaptar el plan de estudios y los métodos didácticos si quieren preparar a sus alumnos para este importante examen.

Después de décadas de este \guillemotleft sistema autopropagante\guillemotright ~centrado en la memorización de conocimientos fácticos, no es fácil para los docentes realizar el cambio. Son pocos los profesores que han experimentado alguna vez la educación científica centrada en el alumno o que estimule el desarrollo de habilidades. Las carreras universitarias pedagógicas ofrecen muy poca orientación más allá de las discusiones teóricas, dado que los futuros profesores generalmente se preparan en clases magistrales tradicionales~\cite{girep,BARRETTO2015}. La mayoría de los educadores están de acuerdo con que la memorización de conocimiento fáctico no es suficiente para preparar a los alumnos a tener una carrera lucrativa o una vida gratificante con participación social consciente. Así y todo, muchos se resisten a desviarse de la conocida \guillemotleft clase teórica\guillemotright ~para comenzar a implementar experimentos prácticos. Las razones más mencionadas están resumidas a continuación:

\begin{figure}
	\centering
	\includegraphics[width=0.66\linewidth,clip=true]{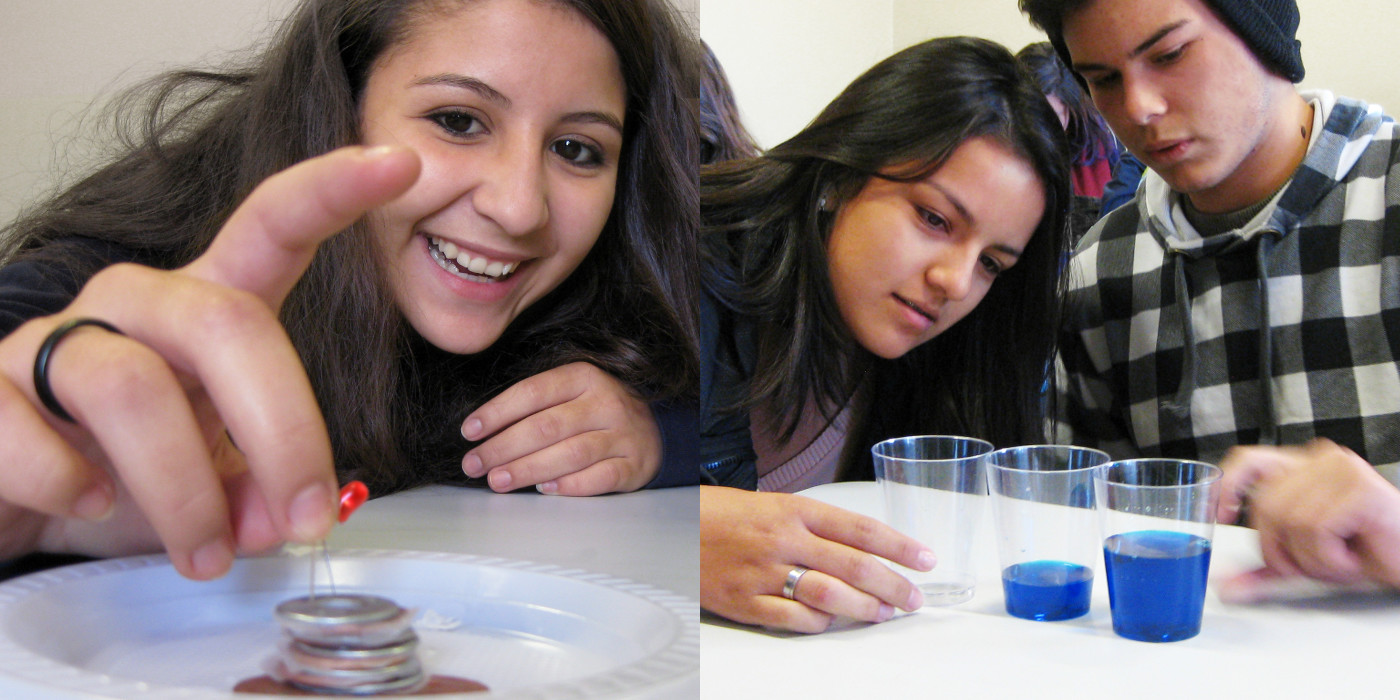}
	\caption[students]{Los experimentos prácticos no solo despiertan el interés de los alumnos por la ciencia, sino que son un componente esencial para la educación científica que estimula el desarrollo de habilidades~\cite{kyle}.}
	\label{fig:studens}
\end{figure}

\begin{itemize}
\item {\bf Falta de tiempo:} como en la mayoría de los sistemas educativos, los docentes de ciencias brasileños tienen que apretar un currículo escolar sobrecargado en un período de tiempo corto. Como ejemplo cuantitativo, los alumnos que se graduaron en 2016 en la escuela técnica secundaria IFAL —una institución del gobierno federal— deberían haber tenido 267 horas de clases de física en 3 años. Debido a exámenes y a tareas administrativas, como verificar la asistencia de cada alumno, tuvieron efectivamente menos de 200 horas de física, sin considerar ausencias por enfermedad y huelgas frecuentes en las instituciones educativas. De esta manera, en menos de 200 horas, los profesores tuvieron que cubrir todos los contenidos de física clásica, inclusive mecánica, óptica, termodinámica y electromagnetismo. Como el enfoque tradicional centrado en el docente permite avanzar rápidamente en los contenidos, es comprensible que bajo estas condiciones los profesores encuentren poco lugar para realizar actividades con los alumnos que demandan mucho tiempo, como el caso de los experimentos prácticos.

\item {\bf Inseguridad y falta de capacitación:} aun los docentes con más experiencia pueden sentirse incómodos estando al frente de una clase en la que los alumnos experimentan de forma autónoma, especialmente si nunca lo han hecho antes. Esta situación tan poco familiar requiere habilidades de manejo del aula diferentes. Además, cuando los alumnos están verdaderamente involucrados en un experimento, surgen preguntas que el profesor tal vez no sepa responder. Admitir no saber una respuesta es esencial para el avance científico, pero es percibida como una actitud opuesta al rol cultural del docente como autoridad. En el caso de profesores con pocos conocimientos de contenido, les resulta más fácil alinear esos contenidos con los libros de texto y centrarse en ejercicios matemáticos: según el Censo Escolar de 2011 en Alagoas, un estado ubicado en la región más pobre del Nordeste de Brasil, son más matemáticos que físicos los que enseñan física. Solo el 20~\% de las clases de física en escuelas públicas fueron impartidas por profesores con estudios terciarios en física (o cursando en aquel momento)~\cite{alagoas}. En otros estados brasileños se registraron problemas similares~\cite{BARRETTO2015}.

\item {\bf Falta de recursos e infraestructura:} este es un problema más imaginario que real. Se cree que los experimentos científicos requieren herramientas sofisticadas y laboratorios especializados. Para proteger esos costos equipos y evitar accidentes, los alumnos solo pueden utilizarlos siguiendo instrucciones detalladas paso a paso, en lugar de seguir su curiosidad científica. Los docentes que buscan alternativas encontrarán en internet una cantidad de experimentos para realizar con materiales domésticos. Sin embargo, estos experimentos son formulados generalmente para entretenimiento y no para aprendizaje. Estas actividades colocan al docente en el papel de \guillemotleft mago\guillemotright , dando más importancia a los efectos visuales o acústicos antes que a la claridad, y requieren mucho tiempo de preparación. Aunque sean considerados seguros bajo la supervisión de un adulto, pueden dejar de serlo si este adulto tiene que cuidar y estar pendiente de 40 o 50 alumnos realizando el experimento a la vez. 
\end{itemize} 

Otra razón por la cual los profesores de ciencia tienden a no incluir experimentos prácticos en sus clases es la costumbre local de separar actividades experimentales de clases teóricas. Especialmente en escuelas privadas, el trabajo de laboratorio a menudo se lleva a cabo bajo la supervisión de educadores especializados y apenas se relaciona con el respectivo contenido teórico enseñado por el profesor de física o química. Pueden pasar semanas hasta que los alumnos que realizaron un experimento debatan con el docente qué fenómenos físicos o reacciones químicas causaron los efectos observados. O tal vez hayan hablado de determinado efecto semanas antes de poder verlo.

\section{Integración de actividades de investigación prácticas en la enseñanza del día a día}
Una iniciativa como Ciensación no va a eliminar estos obstáculos, pero puede ayudar a los docentes a sortearlos. En la práctica, esto quiere decir: convencer a los docentes de que los experimentos cortos —que se pueden conducir en apenas un par de minutos o hasta segundos— pueden generar debates productivos, en los que los alumnos potencian los conocimientos fácticos, profundizan la comprensión y desarrollan habilidades científicas~\cite{Hattie}. Esto se basa en la suposición de que los profesores que tengan buenas experiencias aplicando este enfoque, lo puedan adaptar a su propio estilo de enseñanza para que pase a formar parte de sus prácticas cotidianas en el aula.

Los programas de capacitación docente son un medio eficaz para transmitir esta idea, pero su alcance está limitado por recursos humanos y financieros. Con el propósito de ofrecer este material a un público más amplio, el proyecto Ciensación ha sido implementado como un compilado en línea de Recursos Educativos Abiertos. Los experimentos prácticos están publicados en portugués, español e inglés para que los docentes de toda Latinoamérica puedan tener acceso a ellos. Estos recursos no solo son abiertos en el sentido de estar disponibles gratuitamente bajo una licencia Creative Commons, sino que también con la idea de que cualquier persona puede contribuir enviando su propio experimento (que pasará por un proceso de revisión). Los educadores profesionales necesitan este tipo de libertad editorial para adaptar el material didáctico a las necesidades de sus alumnos. Al compartir el contenido que ellos van creando en el proceso, moldean el crecimiento de Ciensación y ayudan a que se adapte de forma dinámica a los intereses de la comunidad~\cite{Marcos}.

Los experimentos publicados están formulados para que se integren fácilmente a clases regulares en aulas normales. Por lo tanto, la principal preocupación es el tiempo y esfuerzo necesarios para preparar y realizar un experimento. Cuanto menos tiempo tenga que invertir un profesor, más abierto estará para probar una nueva metodología. De esta forma, algunos experimentos muy conocidos fueron modificados para reducir el tiempo de clase utilizado, lo cual muchas veces llevó a simplificarlos.

\begin{figure}
	\centering
	\includegraphics[width=0.66\linewidth,clip=true]{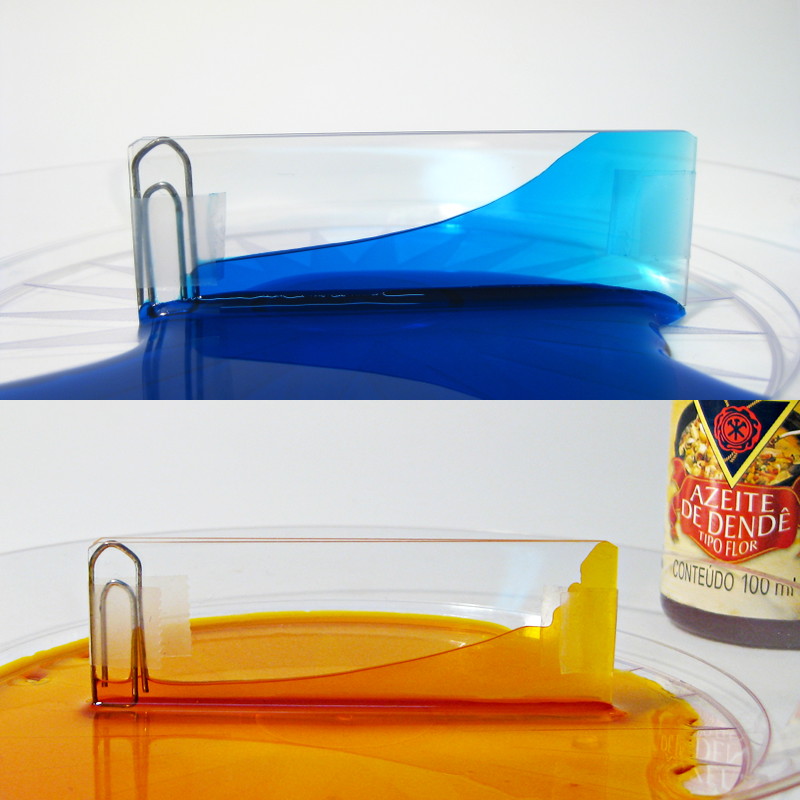}
	\caption[capillary]{Simplificar experimentos a menudo significa trabajar con resultados cualitativos antes que cuantitativos. En este ejemplo, la capilaridad se estudia mediante la comparación del efecto en agua (azul) y aceite vegetal (naranja) con una cuña de diferentes tamaños formada con dos portaobjetos.
}
	\label{fig:capillary}
\end{figure}

En muchos casos, esto significa trabajar con resultados cualitativos obtenidos mediante observación directa, en lugar de mediciones cuantitativas. Además de reducir la complejidad de la actividad, concentrarse en resultados cualitativos también ayuda a alcanzar el objetivo real de los experimentos de Ciensación: involucrar a los alumnos en un debate basado en evidencias sobre la interpretación de sus observaciones. Para insertar los experimentos en un contexto de aprendizaje significativo, todos ellos fueron creados en torno de una tarea de investigación para los alumnos. Las preguntas o desafíos propuestos en la tarea proporcionan una orientación clara sobre lo que el alumno debería descubrir y aprender del experimento. A veces, algunas preguntas aparentemente sencillas, como \guillemotleft ¿Las dos puntas de un imán son iguales?\guillemotright , pueden llevar a extensos debates donde los docentes se enfrentan al desafío de encontrar el delgado límite entre contener el debate y permitir a los alumnos desarrollar sus ideas. Dominar ese equilibrio es muy importante para que la experiencia de aprendizaje sea de calidad y totalmente eficiente. Para evitar frustraciones tanto de los alumnos como de los profesores, Ciensación ofrece \guillemotleft preguntas orientadoras\guillemotright , para dividir la tarea de investigación en pasos más cortos y que, si son usadas con estilo socrático, pueden hacer avanzar la discusión hacia la dirección deseada. Además, estas preguntas sirven para que los docentes puedan hacer la transición de un experimento a su respectivo marco teórico sin perder el impulso motivador de la actividad. 

Los experimentos de Ciensación también contienen una explicación concisa del fenómeno observado, que funciona como punto de partida para investigaciones más detalladas. La intención no es reemplazar libros de texto, sino reducir el tiempo de preparación y ayudar a que los docentes con menos experiencia puedan guiar el debate en clase sintiéndose seguros.

Los materiales utilizados en estos experimentos prácticos deben ser seleccionados con mucho cuidado. Ningún docente responsable aceptaría realizar actividades que puedan poner en riesgo la salud de sus alumnos, y tampoco se espera que compren equipos costosos que la escuela no tiene condición de proveer. Por lo tanto, los experimentos de Ciensación deben basarse en materiales seguros, accesibles y baratos. Cuando es posible, son objetos domésticos, preferentemente que los alumnos tengan a mano, como libros, lápices o monedas. Por un lado, los materiales simples, genéricos y versátiles cuentan con la ventaja de permitir que los alumnos experimenten libremente y dejen llevar su curiosidad científica hacia direcciones inesperadas~\cite{photonicsExplorer}. Por otro lado, las variaciones en el equipo utilizado pueden generar confusión entre los alumnos, y los aparatos que requieren muchos ajustes pueden distraer la atención de la clase hacia la preparación y no hacia el fenómeno real, además de consumir tiempo de clase. Por lo tanto, antes de publicar un experimento en la página, se realizan pruebas de diferentes formas y disposiciones para garantizar que sea seguro y encontrar el equilibrio preciso entre simplicidad, flexibilidad y costo de los materiales.

\begin{figure}
	\centering
	\includegraphics[width=0.5\linewidth,clip=true]{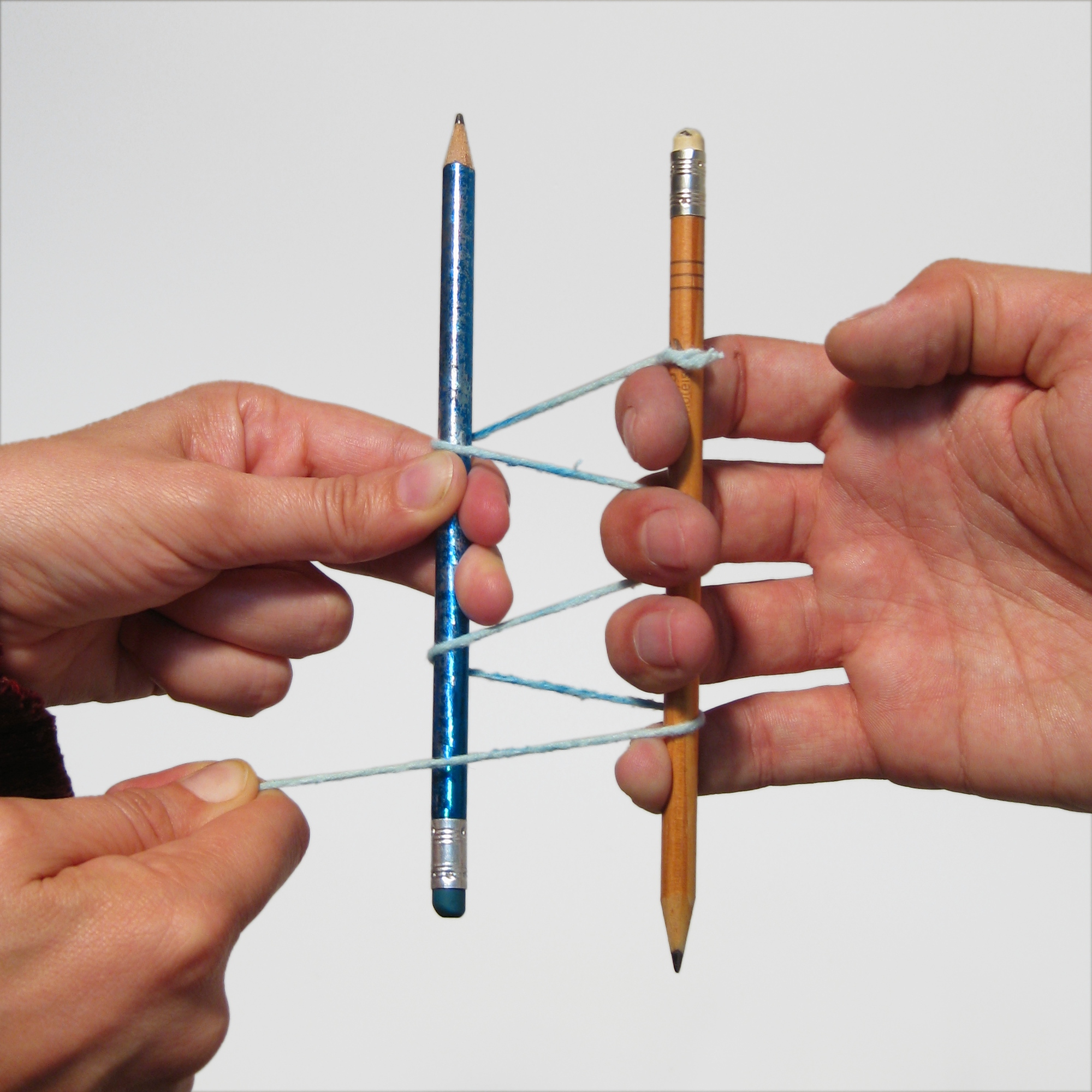}
	\caption[pulley]{Para adquirir una noción, o una comprensión intuitiva, de sistemas mecánicos como poleas, los alumnos necesitan sentir las fuerzas involucradas con sus propios sentidos. Como se muestra en el ejemplo de la polea, podemos lograrlo con medios sencillos que los alumnos suelen tener a mano.
}
	\label{fig:pulley}
\end{figure}

\section{Sentir la ciencia}
Además de facilitar la integración de los experimentos de Ciensación en la enseñanza regular, las actividades fueron formuladas para estimular el desarrollo de habilidades científicas, tales como pensamiento crítico, argumentación basada en evidencias e investigación experimental. Asimismo, tienen por objetivo permitir que los alumnos sientan la ciencia, idea que le dio el nombre al proyecto: Ciensación es un neologismo uniendo las palabras \guillemotleft ciencia\guillemotright  ~y \guillemotleft sensación\guillemotright ~(la página web en inglés y en portugués mantienen el mismo neologismo: sciensation.org y ciensacao.org).

Obviamente, ningún individuo responsable permitirá que los sentimientos subjetivos interfieran a la hora de analizar evidencia científica. Sin embargo, comúnmente se asocia la ciencia con conocimientos estructurados, lo que muchas veces (y sin querer) implica que la ciencia es principalmente una recopilación de conocimientos. Esto reduce el concepto de ciencia —que no se trata tanto de conocimientos, sino más de la habilidad, la metodología y el placer de descubrir— a un simple conjunto de resultados. De la misma forma, podríamos reducir el concepto de arte a \guillemotleft cosas que hay en los museos\guillemotright .

Las tareas de investigación, alrededor de las cuales se crearon los experimentos de Ciensación, invitan a los alumnos a ser verdaderos científicos, en lugar de simplemente reproducir resultados ya conocidos y confirmar lo que dicen los libros de texto. Dar a los alumnos algunos minutos para resolver esa tarea en pequeños grupos y de forma autónoma permite no solo que sientan la emoción de descubrir, sino que también sientan la ciencia como una actividad creativa, un arte que pueden dominar, y que no la vean como un privilegio de una elite llamada \guillemotleft científicos\guillemotright .

La idea de sentir la ciencia también se extiende a la de adquirir una comprensión intuitiva de sistemas físicos o químicos y su comportamiento. En términos cognitivos, es muy diferente calcular la frecuencia de resonancia de un sistema teórico masa-resorte que hacer oscilar un peso en un resorte sosteniéndolo con tu propia mano~\cite{expSprings}. Las leyes de la física suelen tener más \guillemotleft sentido\guillemotright ~si el estudiante puede relacionarlas con una sensación táctil, visual y acústica.  

\begin{figure}
	\centering
	\includegraphics[width=0.5\linewidth,clip=true]{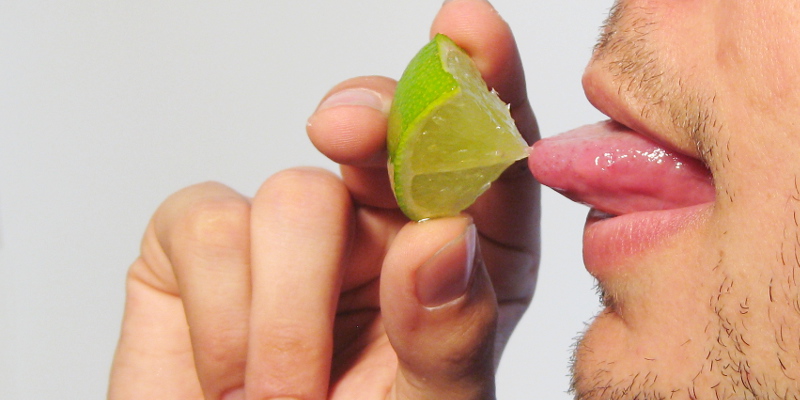}
	\caption[tongue]{Escudriñar mediante experimentos afirmaciones de los libros didácticos, como el mapa de la lengua (que afirma que los diferentes sabores se detectan en distintas regiones de la lengua), ayuda a que los alumnos elaboren evidencias basados en su autonomía intelectual. 
}
	\label{fig:tongue}
\end{figure}

De forma similar, los alumnos pueden aprender grandes lecciones sobre la naturaleza de la ciencia cuando descubran contradicciones entre los libros didácticos y sus propios resultados experimentales. Los errores de algunos libros —como afirmar que los diferentes sabores se detectan en distintas regiones de la lengua~\cite{expTongueMap} o representar el polo norte magnético en el hemisferio norte~\cite{expNorth}~— son muy buenas oportunidades para enseñar la importancia de buscar evidencia científica en lugar de confiar en las autoridades sin sentido crítico. Observaciones contrarias al sentido común, como la atracción de polos magnéticos iguales si se coloca una moneda entre ellos~\cite{expMagnets}, son a menudo más eficientes para perfeccionar las habilidades de investigación de los alumnos que los experimentos con resultados predecibles. Algunos experimentos de Ciensación hasta cuestionan el propio método científico~\cite{expScientificMethod}. Para demostrar que la opinión de la mayoría no equivale a la evidencia científica, en un experimento se pide a los alumnos que observen el movimiento de las manchas de un láser en una pared mientras mueven la cabeza~\cite{expSpeckle}. Va a depender de la vista de cada observador si las manchas se mueven en la misma dirección o en la dirección contraria a la cabeza (una persona con hipermetropía verá las manchas moverse en la misma dirección que su cabeza, mientras que un observador miope las verá moverse en la dirección contraria). Sin embargo, el profesor —engañando a sus alumnos por un momento— presiona a la clase a decidirse por una sola dirección y se opone a todos los que opinen diferente, para luego revelar la verdadera naturaleza del experimento. Si algunos alumnos comienzan a dudar de sus propias observaciones o se rinden ante la presión de los compañeros, toda la clase aprenderá lo que es un obstáculo crítico en los desafíos científicos.

\section{Impacto}
Para descubrir si es posible alcanzar las ambiciosas expectativas que el proyecto propone, Ciensación está siendo probado en cursos de capacitación docente en la Universidad Federal de Paraná (Brasil) y en la Universidad Federal de Fronteira Sul (Brasil). En un taller que se realizó recientemente en la Universidad Federal de Rio Grande do Sul, se reunieron investigadores educativos, profesores de ciencia y alumnos para debatir de qué forma puede difundirse el proyecto Ciensación para que llegue a más docentes y así puedan interactuar de manera más eficiente con la plataforma y entre ellos. 
 
Mientras tanto, docentes de todo el mundo están ingresando a la página web, leyendo los experimentos y dejando comentarios de agradecimientos desde países tan diversos como Gana y Alemania. En Fiyi y en las Islas Salomón, donde el acceso a internet es limitado, se ha distribuido a los docentes una versión offline de los experimentos en tarjetas de memoria. Las devoluciones positivas y el creciente número de visitantes de la página demuestran que los primeros pasos del proyecto Ciensación realmente están contribuyendo a crear una cultura de experimentos prácticos cortos y una educación científica que estimule el desarrollo de habilidades, mucho más allá del público previsto de docentes de Latinoamérica.

\section{Bibliografía}
\bibliography{CSC4IOP}
\bibliographystyle{IEEEtran}
\end{document}


\title[Ciênsação: despertando um sentimento por ciências]{Ciênsação: despertando um sentimento por ciências}

\author{Marcos Henrique Abreu de Oliveira$^1$ and Robert Fischer$^2$}
\address{$^1$Instituto Federal de Alagoas, Maceió, AL, Brazil\\$^2$Universidade Tecnológica Federal do Paraná, Curitiba, PR, Brazil}
\ead{r@robertfischer.eu}

\begin{abstract}
Ciênsação, uma coleção aberta de experimentos práticos, foi desenvolvida para convencer a professores da América Latina que ciência é melhor entendida quando manuseada. Permitir que os estudantes experienciem autonomamente, em pequenos grupos, pode ser desafiador para os educadores desses países. Analisamos as razões pelas quais os professores hesitam em aplicar experimentos práticos em sala de aula, e discutimos como Ciênsação foi elaborada para superar tais obstáculos. As atividades resultantes de pesquisa foram desenhadas, especificamente, para serem facilmente integradas no ensino formal de ciência na escola, afim de promover habilidades de pesquisa e, o que é mais importante, permitir aos estudantes entender a ciência como uma atividade criativa e envolvente. \textit{ [Tradução do original inglês ``Ciênsação: gaining a feeling for sciences'']} 

\end{abstract}


\section{Experiências práticas em escolas brasileiras}
Quando Richard P. Feyman, um entusiástico professor de física, visitou o Brasil em 1951, ele viu para seu deleite ``estudantes da escola primária em livrarias comprando livros de Física''~\cite{feynman}. No entanto, depois de ensinar um semestre em uma universidade no Rio de Janeiro, ele chegou à conclusão de que ``ciência alguma era ensinada no Brasil''. Em suas memórias ele escreveu: ``depois de muita investigação, eu finalmente entendi que os estudantes memorizavam tudo, mas não sabiam o que de fato significava''.  Enquanto os alunos podiam ``recitar, palavra por palavra''  até física avançada, eles não eram capazes de relacioná-la aos problemas do mundo real, ou aplicá-la em experimentos simples. Feynman se referiu a esse tema em uma palestra que ele proferiu no fim do semestre: ``Finalmente, eu disse que não conseguia entender como alguém poderia ser educado por um sistema que se auto propaga, no qual pessoas passam em provas, e ensinam a outras a passar em provas, sem que ninguém saiba coisa alguma.''~\cite{feynman}

Nestes 65 anos da visita de Feynman ao Brasil, o sistema educacional brasileiro fez progressos significativos. Hoje, as instituições educacionais dão mais ênfase à compreensão e à evidência experimental. Os esforços feitos para se discutir fenômenos naturais em um contexto significativo e relevante têm iniciado uma transformação no currículo e em livros escolares, que começam a chegar à sala de aula. Entretanto, experimentos práticos, assim como os defendidos por Feynman, ainda são raros nas escolas brasileiras. Tais atividades centradas no aluno, que em pequenos grupos exploram fenômenos naturais com suas próprias mãos e mentes, não são meramente aplicadas por diversão, ou para aumentar o interesse dos jovens por ciência, mas traz contribuições cruciais para o desenvolvimento dos aprendizes, promovendo o desenvolvimento de habilidades e a autonomia intelectual baseada na evidência. Para entender porque professores no Brasil hesitam em fazer uso deste método didático comprovado~\cite{Hattie,stohr}, é importante olhar para a situação local e dificuldades enfrentadas por eles em sala de aula. Por muitos anos, o projeto Ciênsação tem estudado o ambiente de aprendizagem com respeito às políticas educacionais, a formação de professores e práticas de ensino diárias nas escolas secundárias e primárias, antes de fazer uma abordagem aos obstáculos comuns e conquistar os professores para experimentos práticos em sala de aula.

Já no ano de 1997, o Ministério da Educação publicou no seu currículo nacional~\cite{PCN} a recomendação: ``o experimento se torna mais importante quanto mais os alunos participam na  confecção de seu guia ou protocolo, realizam por si mesmos as ações sobre os materiais e discutem os resultados.'' Poucos professores seguiram esse conselho na época. Recentemente, entretanto, as políticas educacionais aumentaram a ênfase na importância de desenvolver habilidades, como pensamento crítico e analítico, trabalho em equipe e comunicação efetiva~\cite{bruns}, que estão intimamente relacionadas aos experimentos práticos~\cite{kyle,stohr}. Como exemplo, o ENEM, um exame de admissão nacional para universidades públicas, foi remodelado para testar habilidades e competências. A memorização de conhecimento factual ainda tem um papel importante, mas para responder a questões de ciência agora é necessária uma compreensão real. Consequentemente, os professores precisam adaptar seus currículos e métodos de ensino, caso queiram preparar seus alunos para esse importante exame nacional.

Depois de décadas neste ``sistema auto-propagante'' que focalizava a atenção na memorização de conhecimentos factuais, não é fácil para os professores realizar essa mudança. Poucos deles vivenciaram uma educação centrada no estudante, que desenvolvesse habilidades em ciência. As licenciaturas nas universidades oferecem pouca orientação além das discussões teóricas, já que os futuros professores são normalmente preparados em seminários e palestras tradicionais~\cite{girep,BARRETTO2015}. A maioria dos educadores prontamente concordará que só a memorização de fatos não é suficiente para preparar o aluno para uma carreira de sucesso, ou mesmo a realização pessoal com participação social consciente. Além disso, muitos se recusam a abandonar seu `cospe-giz' diário para incluir experimentos práticos em suas aulas. As razões comumente citadas podem ser assim resumidas:

\begin{figure}
	\centering
	\includegraphics[width=0.66\linewidth,clip=true]{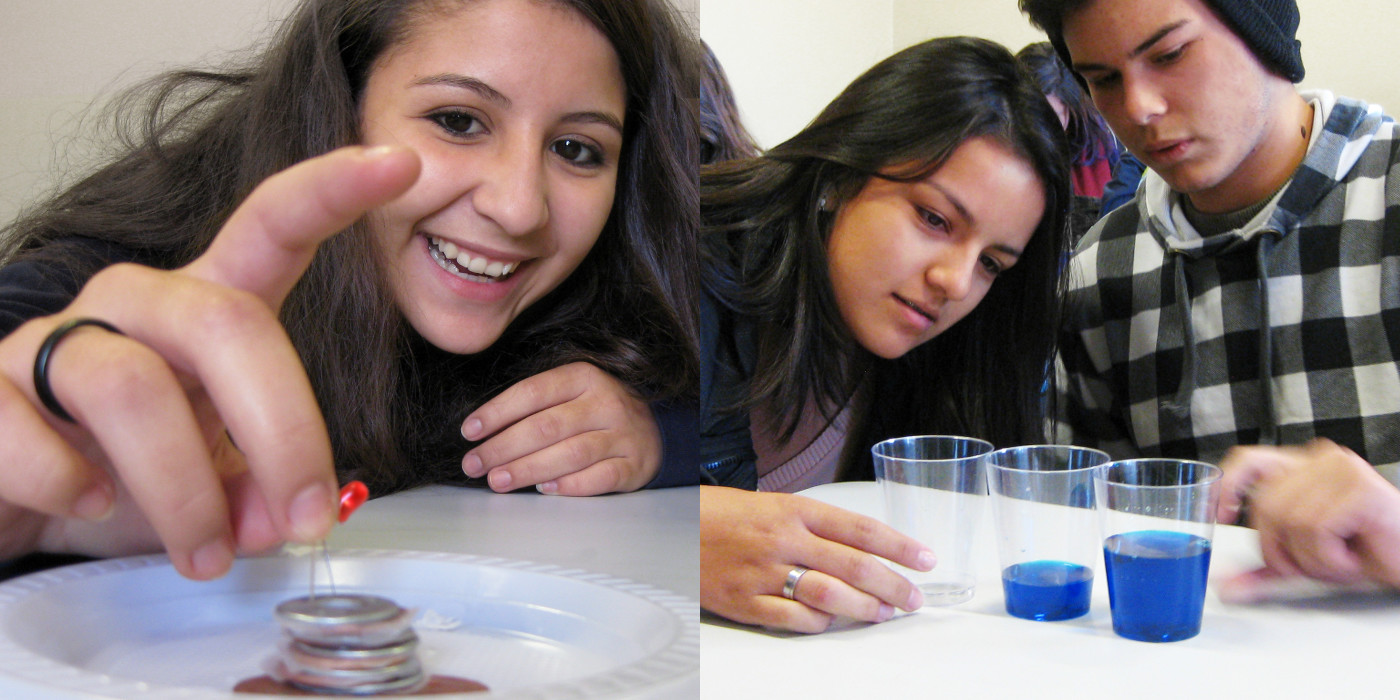}
	\caption[students]{Experimentos práticos não apenas despertam o interesse do estudante por ciência, mas são elementos essenciais para promover habilidades na educação científica~\cite{kyle}.}
	\label{fig:studens}
\end{figure}

\begin{itemize}
\item {\bf Falta de tempo:} como na maioria dos sistemas educacionais, os professores de ciência brasileiros têm que comprimir um currículo sobrecarregado em um pequeno intervalo de tempo. Para dar um exemplo quantitativo, os estudantes dos cursos técnicos secundários em 1916 no IFAL, uma instituição administrada pelo governo brasileiro, tinham que assistir 267 horas de física em três anos. Devido a testes e atividades administrativas, tais como verificar a presença de cada aluno, eles efetivamente ofereciam menos que 200 horas de física, sem levar em conta ausências devido a problemas de saúde e frequentes greves nas instituições educacionais. Sendo assim, em menos de 200 horas, os professores tinham que cobrir o conteúdo de física clássica, incluindo mecânica, óptica, termodinâmica e eletromagnetismo. Como a abordagem tradicional de ensino centrada no professor permite avançar mais rápido nos conteúdos, é compreensível que, sob tais condições, os professores encontrem pouco tempo para realizar demoradas atividades com os estudantes, como experimentos práticos..

\item {\bf Insegurança e falta de treinamento:} até os professores com mais experiência podem se sentir desconfortáveis ao supervisionar um grupo de experimentação autônoma – especialmente se jamais fizeram isso antes.  Tal situação, não-familiar, requer diferentes habilidades de manejo de turma. Além do mais, quando os alunos estão realmente engajados em um experimento, podem surgir perguntas e talvez o professor não esteja preparado para respondê-las. Embora admitir que não se sabe a resposta seja essencial para o progresso científico, esta percepção está em desacordo com o papel cultural do professor como autoridade. Para professores com conhecimento limitado do assunto é preferível, nestas circunstâncias, alinhar o conteúdo abordado ao livro adotado. E, de acordo com o Censo Escolar oficial de Alagoas em 2011, um dos estados mais pobres do nordeste brasileiro, mais matemáticos estavam ensinando física do que propriamente físicos. Apenas 20~\% das aulas de física nas escolas públicas eram ministradas por professores (graduados ou graduandos) com formação superior na área~\cite{alagoas}. Problemas similares foram reportados em outros estados brasileiros~\cite{BARRETTO2015}.

\item {\bf Falta de recursos e infraestrutura:} Este é mais um problema imaginário do que real. Acredita-se que experimentos científicos requerem equipamentos sofisticados e laboratórios especializados. Afim de proteger esses equipamentos caros de acidentes, permite-se aos estudantes operá-los apenas seguindo instruções detalhadas, em vez de permitir que sigam sua curiosidade científica. Professores que procuram alternativa para isso encontrarão uma variedade de recursos online que propõem experimentos com materiais domésticos. Esses experimentos, entretanto, são normalmente pensados para entretenimento, não para o ensino. Levando o professor a assumir um papel de  'mágico', que valoriza impressionantes efeitos visuais ou acústicos em detrimento da clareza, além de demandar muito tempo de preparação. Mesmo que eles sejam considerados seguros, sob a supervisão de um adulto, isso não é bem verdade se, o adulto, tiver que observar de 40 a 50 alunos ao mesmo tempo.
\end{itemize} 

Uma outra razão pela qual os professores de ciência tendem a não incluir experimentos práticos em suas aulas é o costume local de separar atividades práticas de aulas teóricas. Particularmente em escolas privadas, o trabalho no laboratório é frequentemente conduzido sob a supervisão de um educador especialista e ligeiramente alinhado com a perspectiva do conteúdo teórico ensinado pelos professores de física e química. Podem se passar semanas até que os alunos que participam do experimento possam discutir com seus professores, quais fenômenos físicos ou reações químicas causaram os efeitos observados. Ou talvez eles tenham discutido sobre um determinado efeito semanas antes de observá-lo.

\section{Integração de atividades práticas de pesquisa no ensino diário}
Uma iniciativa como Ciênsação não será capaz de remover todos esses obstáculos, mas poderá ajudar aos professores a superá-los. Na prática, significa convencer aos professores que pequenos experimentos – que levam só alguns minutos ou mesmo segundos para serem conduzidos – podem, de forma suave, levar a produtivas discussões em sala, nas quais os alunos simultaneamente avançam no conhecimento factual, aprofundam seu entendimento e desenvolvem suas habilidades científicas~\cite{Hattie}. Isto se baseia na suposição de que os professores que tiveram uma experiência positiva com essa abordagem, irão adaptá-la ao seu estilo de ensino pessoal e torna-la parte de sua prática de ensino diária.

Os programas de treinamento para professores são um método efetivo para transmitir essa ideia, mas são limitados em seu alcance por recursos financeiros e humanos. Afim de fornecer o material para um público maior, o projeto Ciensação tem sido implementado como uma coleção online de recursos educacionais abertos. Experimentos práticos são publicados nesta coleção em português, espanhol e inglês, de modo que ficam acessíveis a professores em toda a América Latina. Esses recursos são abertos não apenas no sentido de estarem gratuitamente disponíveis sob uma licença Creative Commons, mas também no sentido de que qualquer um pode contribuir submetendo sua própria experiência (que será revisada por outros professores de ciências). Educadores profissionais precisam de tal liberdade editorial para adaptar materiais de curso às necessidades de seus alunos. Ao compartilhar conteúdos que eles criam no processo, eles moldam o desenvolvimento do Ciênsação e ajudam a ajustá-lo dinamicamente aos interesses da comunidade~\cite{Marcos}. 

Os experimentos publicados foram planejados para serem facilmente integrados às aulas regulares em sala de aula. Sendo assim, uma preocupação primordial é com o tempo e esforço requerido para preparar e conduzir o experimento. Quanto menos tempo o professor tiver que investir, mais aberto ele estará para tentar a nova metodologia.  Portanto, alguns experimentos bem conhecidos foram modificados para reduzir o tempo investido de aula, e isso acompanha par-e-passo o processo de simplificação deles.

\begin{figure}
	\centering
	\includegraphics[width=0.66\linewidth,clip=true]{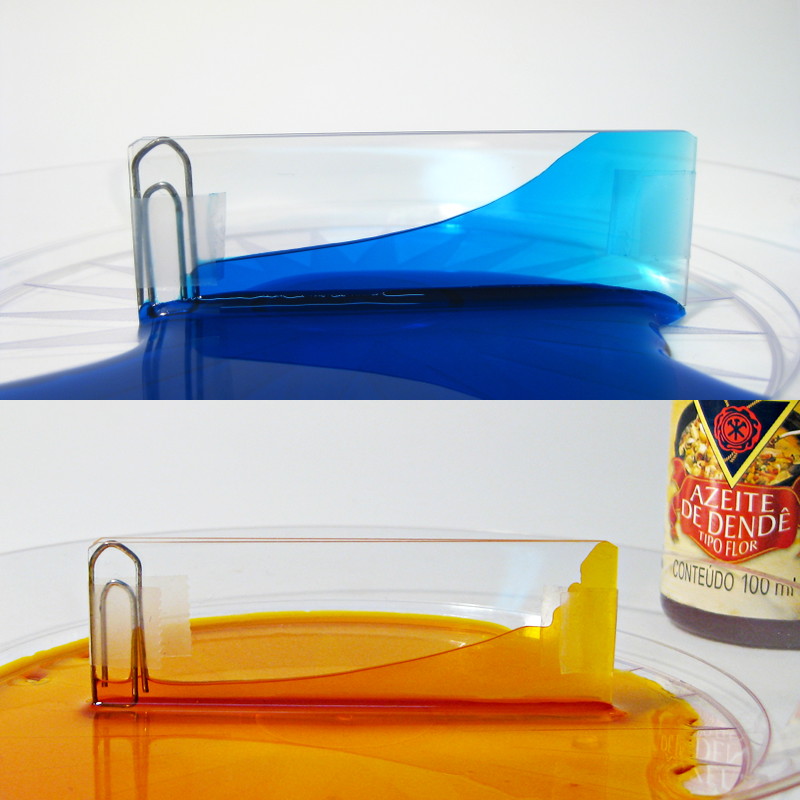}
	\caption[capillary]{Experimento simplificado geralmente significa trabalhar com resultados qualitativos em detrimento dos quantitativos. Neste exemplo, a capilaridade é estudada comparando seus efeitos na água (em azul) e no óleo vegetal (em laranja) para um espaço vazio de diferentes larguras em um recipiente com duas chapas de vidro.
}
	\label{fig:capillary}
\end{figure}

Em muitos casos, isso significa trabalhar com resultados qualitativos em vez de com medidas quantitativas. Além de minimizar a complexidade das atividades, a concentração em resultados qualitativos também serve ao propósito real dos experimentos do Ciênsação, mais especificamente no que se refere em engajar os alunos em discussões baseadas na evidência sobre a interpretação de suas observações. Para incorporar experiências em um contexto de aprendizagem significativa, cada uma é construída em função de uma tarefa de pesquisa específica do aluno. As questões das tarefas ou desafios aos alunos fornecem uma orientação clara, indicando o que o aluno deve encontrar ou aprender com tal experimento. Às vezes, mesmo questões aparentemente simples – como: Os dois terminais de um ímã são diferentes? – podem levar a uma discussão generalizada, o que desafia aos professores a encontrar uma linha tênue entre restringir a discussão e permitir aos alunos a desenvolverem suas ideias. Dominar esse equilíbrio é de grande importância para a eficácia global, bem como para a qualidade da aprendizagem na experiência. Afim de evitar frustração para professores e alunos, Ciênsação oferece 'perguntas orientadoras', o que divide a tarefa da pesquisa do aluno em passos menores, que quando realizados em um estilo socrático, podem avançar as discussões na direção pretendida. Além do mais, essas questões podem ser úteis ao professor para fazer uma transição suave de um experimento para o conteúdo teórico respectivo sem perder o momento motivacional da atividade. 

Os experimentos da Ciênsação também incluem uma explicação concisa do fenômeno observado, que funcionam como ponto de partida para investigações mais detalhadas. A intenção não é substituir livros-textos, mas reduzir o tempo de preparação e ajudar professores menos experientes a lidar com as discussões em sala-de-aula com maior segurança.

Os materiais utilizados em tais experimentos práticos devem ser selecionados com muito cuidado. Nenhum professor responsável aceitaria atividades que poderiam colocar a saúde de seus alunos em risco, nem se pode esperar que eles comprem equipamentos caros, os quais suas escolas não são capazes de fornecer. Consequentemente, os experimentos de Ciênsação precisam ser seguros e contar com materiais de baixo custo. Sempre que possível, são feitos com itens domésticos, preferencialmente objetos que os alunos já têm em mãos, como livros, lápis e moedas. Por um lado, matérias versáteis, genéricos e simples tem a vantagem de permitir aos alunos fazerem experimentos mais livremente, e por outro lado, permite que eles sigam sua curiosidade científica em direções não previstas~\cite{photonicsExplorer}. Por outro lado, muita variação nos equipamentos empregados pode provocar confusão entre os alunos, e dispositivos experimentais que necessitam de muito manuseio desviam a atenção do estudante do fenômeno idealizado na preparação – além de aumentar o custo de tempo de ensino. Antes de publicar um experimento na coleção do {\it site}, ele é testado com uma variedade de configurações para garantir a segurança e achar o equilíbrio certo entre flexibilidade, simplicidade e custos.

\begin{figure}
	\centering
	\includegraphics[width=0.5\linewidth,clip=true]{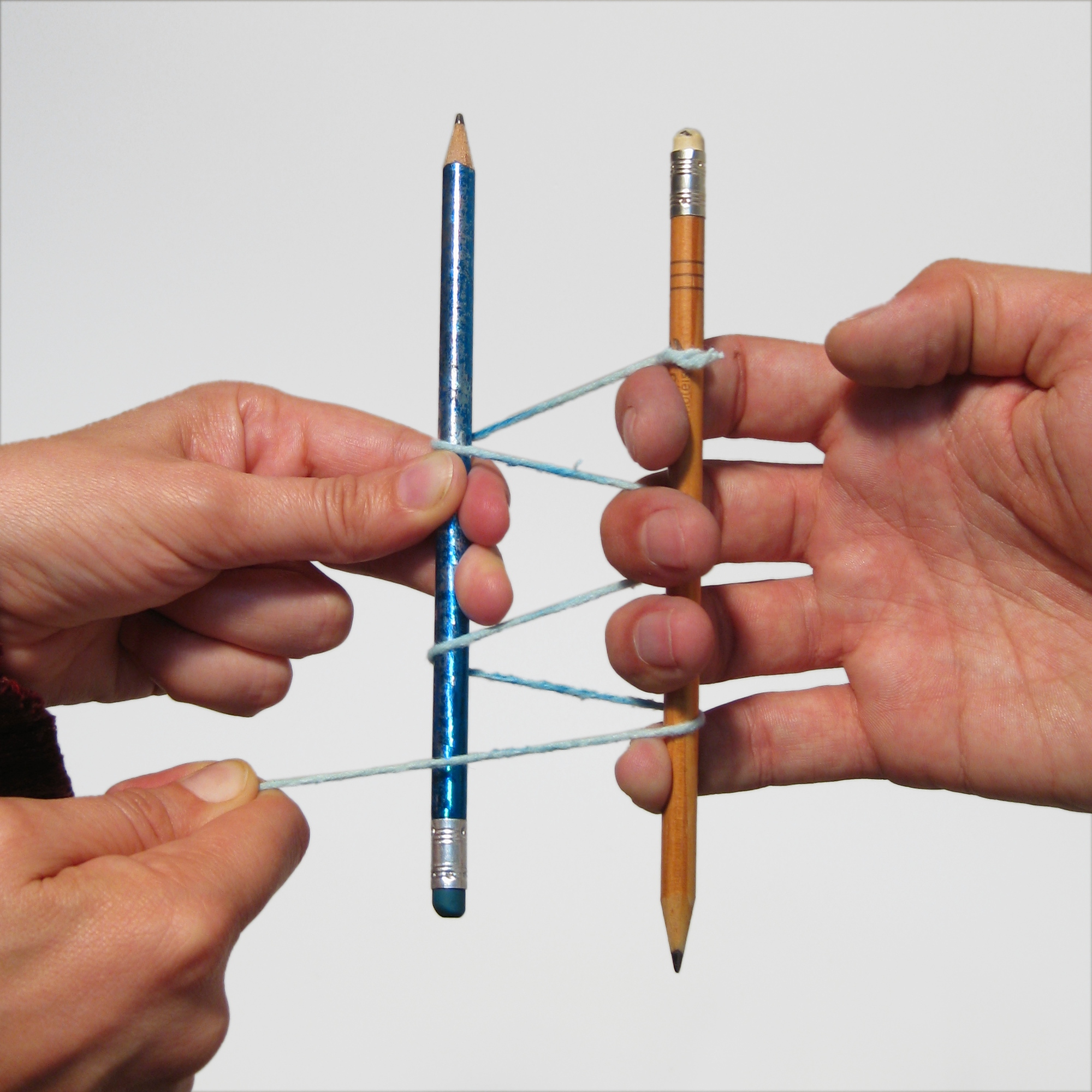}
	\caption[pulley]{Para sentir ou ter uma compreensão intuitiva de sistemas mecânicos, como de uma polia, os alunos precisam experimentar as forças envolvidas com seus próprios sentidos. Como o exemplo da polia demonstra, isso pode se conseguir através de materiais simples, que os alunos já dispõem, usualmente, no seu dia-a-dia.
}
	\label{fig:pulley}
\end{figure}

\section{Despertando um sentimento por ciências}
Além de facilitar aos professores a integração dos experimentos em suas aulas regulares, as atividades de Ciênsação foram especificamente projetadas para desenvolver habilidades científicas, tais como pensamento crítico, argumentação baseada na evidência e pesquisa experimental. Além do mais, eles foram pensados para despertar um sentimento dos alunos pela ciência, o que deu nome ao projeto: Ciênsação. Um neologismo juntando `ciência' e `sensação' (o site da internet em inglês é nominado de acordo com sciensation.org, juntando `science' e `sensation').

Evidentemente, indivíduos responsáveis jamais deixariam sentimentos subjetivos interferirem quando evidências científicas são avaliadas. Entretanto, a associação comum de ciências com fatos sólidos, frequentemente (involuntariamente) implica no entendimento que a ciência é primeiramente uma coleção de fatos conhecidos. Isso reduz o conceito de ciência – que não se trata tanto de fatos, mas de habilidades, metodologia e o prazer de descobri-los – a um simples conjunto de resultados. Da mesma maneira, poder-se-ia reduzir o conceito de arte a ``coisas que se encontram em um museu''.

Essas tarefas de pesquisa, com as quais os experimentos de Ciênsação são construídos, convidam aos alunos a \textit{fazer} ciência ativamente, em vez de reproduzir meramente resultados conhecidos e confirmar afirmações dos livros didáticos. Dar aos alunos alguns minutos para executar tais tarefas de forma autônoma e em pequenos grupos permite que eles não somente experimentem o excitamento da descoberta, mas também experimentem a ciência como uma atividade criativa, como um ofício que eles podem dominar, e não como um privilégio de uma elite chamada de `cientistas'. 

A ideia de despertar um sentimento por ciências também se estende a adquirir um entendimento intuitivo de sistemas físicos e químicos e seus comportamentos. Há uma grande diferença cognitiva entre calcular a frequência de um sistema teórico massa-mola, ou fazer com que um peso oscile em uma mola presa na própria mão~\cite{expSprings}. As leis físicas normalmente fazem mais sentido para um aprendiz quando podem ser relacionadas a suas sensações visuais, acústicas ou táteis.  

\begin{figure}
	\centering
	\includegraphics[width=0.5\linewidth,clip=true]{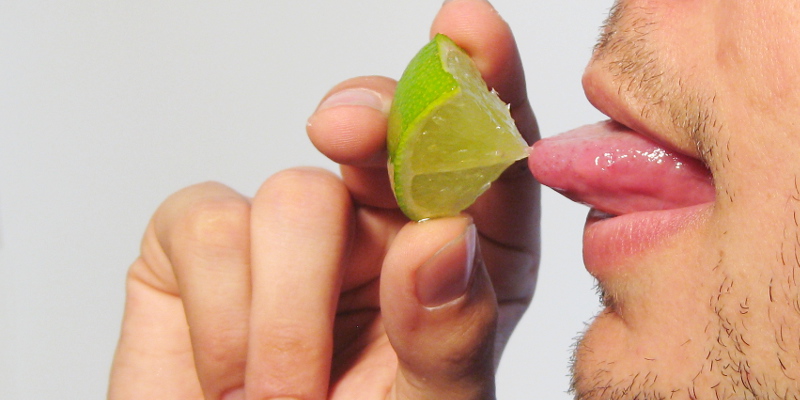}
	\caption[tongue]{Verificar por meio de experimentos práticos as afirmações dos livros didáticos, como o mapa da língua (que afirma que diferentes gostos são percebidos em regiões distintas da língua) pode ajudar aos alunos a desenvolver uma autonomia intelectual baseada na evidência. 
}
	\label{fig:tongue}
\end{figure}

Do mesmo modo, os alunos podem aprender lições importantes sobre a natureza da ciência quando descobrem contradições entre livros didáticos e seus próprios resultados experimentais. Erros comuns em livros textos – tal como afirmar que diferentes gostos são sentidos em regiões distintas da língua ~\cite{expTongueMap}, ou representar o polo norte magnético no hemisfério norte ~\cite{expNorth}~– fornecem oportunidades valiosas para ensinar a importância da procura da evidência científica em vez de, sem nenhum senso crítico, confiar em autoridades. Observações contra-intuitivas, tais como a atração de polos magnéticos iguais se uma moeda for colocada entre eles~\cite{expMagnets}, são frequentemente mais efetivas para aprimorar as habilidades de pesquisa dos alunos que experimentos com resultados previsíveis. Alguns experimentos de Ciênsação até discutem o próprio método científico~\cite{expScientificMethod}. Para demonstrar que a opinião da maioria não é igual a evidência científica, um experimento pede aos alunos que observem as manchas provocadas por um laser em uma parede enquanto movem suas cabeças~\cite{expSpeckle}. Vai depender da visão individual de cada observador se o padrão de manchas se move na mesma direção ou na direção oposta à cabeça (observadores com hipermetropia verão o movimento das manchas na direção do movimento da cabeça, enquanto os míopes as verão se movendo em sentido oposto). No entanto, o professor – enganando seus alunos temporariamente – pressiona a turma a decidir-se por uma direção e contradiz opiniões diferentes, antes de revelar a verdadeira natureza do experimento. Se alguns alunos começarem a duvidar de sua própria observação, ou mesmo ceder à pressão do grupo, a turma toda conhecerá uma armadilha crítica do desafio científico.

\section{Impacto}
Para descobrir se é possível atingir as ambiciosas expectativas que o projeto propõe, Ciênsação está sendo testado em cursos de treinamento de professores na Universidade Federal do Paraná (Brasil) e na Universidade Federal da Fronteira Sul (Santa Catarina, Brasil). Em uma oficina realizada recentemente na Universidade Federal do Rio Grande do Sul, reuniram-se pesquisadores educacionais, professores de ciências e estudantes para discutir como Ciênsação pode ser estendido para atingir mais professores e permitir que eles possam interagir mais eficientemente com a plataforma e entre eles. 
Enquanto isso, experimentos em uma coleção online são acessados por professores de todo o mundo, e comentários apreciativos foram recebidos por países tão diversos como Alemanha e Gana. Uma versão off-line dos experimentos foi distribuída em pendrives a professores das Ilhas Fiji e das Ilhas Salomão, onde o acesso à Internet é limitado. Este feedback encorajador e o grande número de visitantes online indicam que os primeiros passos do projeto Ciênsação de fato contribuem para promover uma cultura de pequenos experimentos práticos de uma educação científica que promove habilidades – o que vai bem além do público-alvo que eram professores da América Latina. 

\section{Bibliografia}
\bibliography{CSC4IOP}
\bibliographystyle{IEEEtran}


\title[Ciênsação: Gaining a feeling for sciences]{Ciênsação: Gaining a feeling for sciences}

\author{Marcos Henrique Abreu de Oliveira$^1$ and Robert Fischer$^2$}
\address{$^1$Instituto Federal de Alagoas, Maceió, AL, Brazil\\$^2$Universidade Tecnológica Federal do Paraná, Curitiba, PR, Brazil}
\ead{r@robertfischer.eu}

\begin{abstract}
Ciênsação, an open online repository for hands-on experiments, has been developed to convince teachers in Latin America that science is best experienced first hand. Permitting students to experiment autonomously in small groups can be a challenging endeavour for educators in these countries. We analyse the reasons that make teachers hesitate to apply hands-on experiments in class, and discuss how Ciênsação was implemented to overcome these obstacles. The resulting student research activities are specifically designed to be easily integrated in formal science education at school, to foster research skills and, most importantly, to let young people get to know science as an engaging creative activity.

\end{abstract}


\section{Hands-on experiments at Brazilian schools}
When Richard P. Feynman, an enthusiastic physics teacher, visited Brazil in 1951, he saw to his delight ``elementary school kids in bookstores, buying physics books''~\cite{feynman}. Yet after teaching a semester at a university in Rio de Janeiro, he came to the conclusion that ``no science is being taught in Brazil''. In his memoirs, he wrote: ``After a lot of investigation, I finally figured out that the students had memorized everything, but they didn’t know what anything meant.''  While students could ``recite, word for word''  even advanced physics, they were not able to relate them to real-world problems, nor could they apply them in simple experiments. Feynman addressed the issue in a talk he gave at the end of the semester: ``Finally, I said that I couldn’t see how anyone could be educated by this self-propagating system in which people pass exams, and teach others to pass exams, but nobody knows anything.''~\cite{feynman}

In the 65 years since Feynman's visit, Brazil's educational system has made significant improvements. Today, educational institutions put more emphasis on understanding and experimental evidence. Efforts to set the discussion of natural phenomena in a meaningful and relevant context, have begun to transform curricula, school books and started to arrive in class rooms. Hands-on experiments, however, as Feynman had advocated them, are still rarely seen at Brazilian schools. Such student-centred activities, in which small groups explore natural phenomena with their own hands and minds, are not merely applied `for fun', or to raise young peoples interest in sciences, but make crucial contributions to the learners development by fostering skills and evidence based intellectual autonomy. To understand why teachers in Brazil hesitate to make use of this proven didactic method~\cite{Hattie,stohr}, it is important to look at the local situation and difficulties faced by teachers when experimenting in class. Over several years, the project Ciênsação has studied this learning environment with regards to educational policies, teacher training and day-to-day teaching practice at primary and secondary schools, before developing an approach to address common obstacles and win over teachers for hands-on experiments in class. 

Already in 1997, the Ministry of Education published in its national curriculum ~\cite{PCN} the recommendation: ``An experiment becomes more relevant to students when they participate in planning it, conduct it themselves, handle the equipment themselves, and then discuss the results.''  Few teachers followed the advice at the time. In recent years, however, educational policies increasingly stress the importance of fostering skills, like critical and analytic thinking, team work and effective communication ~\cite{bruns}, which are closely related to hands-on experiments~\cite{kyle,stohr}. As a case in point, the ENEM, a nationwide entrance exam for public universities, has been remodelled to test skills and competences. Memorised fact knowledge still plays an important role, but answering science questions now requires real understanding. Consequently, teachers need to adapt their syllabus and didactic methods if they want to prepare their students for this important exam.

After decades in this ``self-propagating system'' of being focused on memorised fact knowledge, this is not an easy change for teachers to make. Few of them have ever experienced student-centred or skill fostering science education themselves. University courses on pedagogy offer little orientation beyond theoretical discussions, since future teachers are usually prepared in classical ex-cathedra lectures~\cite{girep,BARRETTO2015}. Most educators will readily agree that memorised fact knowledge alone is insufficient to prepare students for lucrative careers or fulfilling lives with informed social participation. And yet, many hesitate to deviate from their familiar `chalk talk' in order to include hands-on experiments. Commonly cited reasons can be summarised as follows:

\begin{figure}
	\centering
	\includegraphics[width=0.66\linewidth,clip=true]{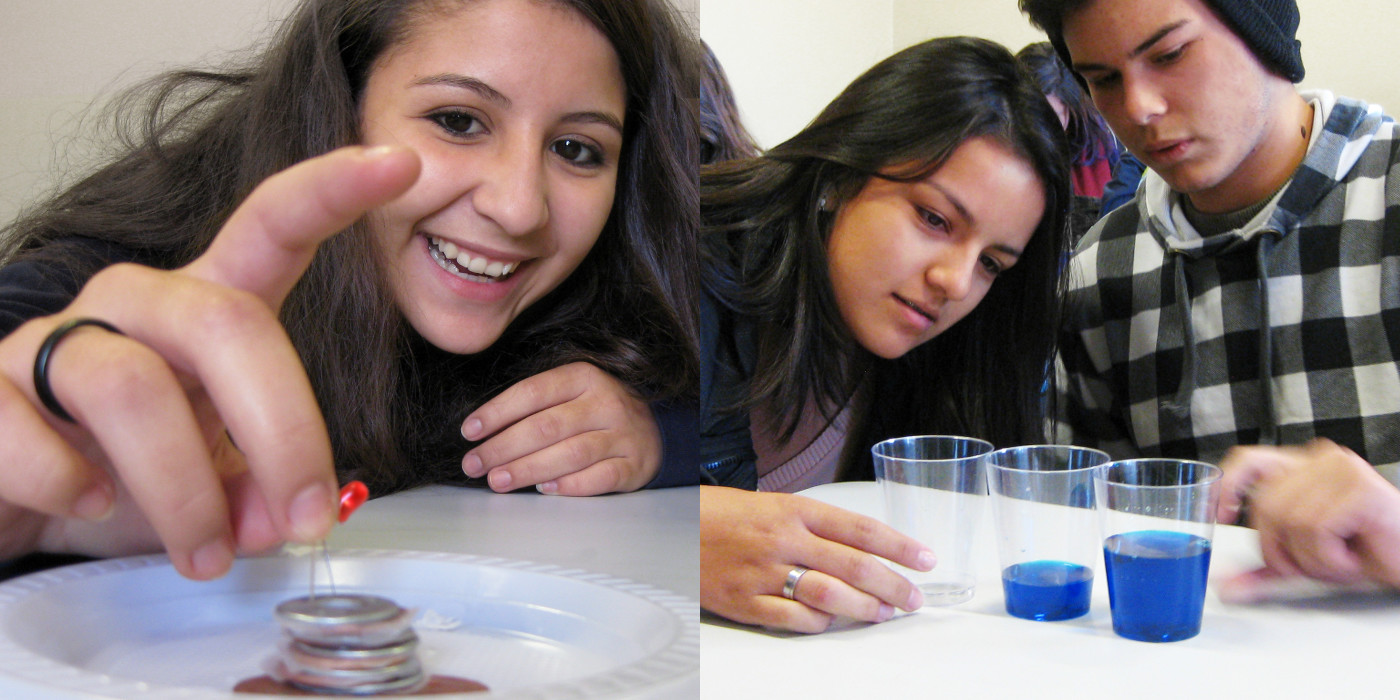}
	\caption[students]{Hands-on experiment not only raise student's interest in sciences, but are an essential component in skill fostering science education.~\cite{kyle}.}
	\label{fig:studens}
\end{figure}

\begin{itemize}
\item {\bf Lack of time:} Like in most educational systems, Brazilian science teachers have to squeeze an overloaded curriculum into a small time frame. To give a quantitative example, students graduating in 2016 from the secondary technical course at IFAL, an institution run by the federal government, should have attended 267 hours of physics over 3 years. Due to exams and administrative tasks, such as verifying the presence of each student, they effectively were offered less than 200 hours of physics, without taking into account sickness related absence and frequent strikes at educational institutions. Thus, in less than 200 hours, teachers have to cover the complete content of classical physics, including mechanics, optics, thermodynamics, and electromagnetism. Since the traditional teacher-centred approach allows to advance faster through the material, it is understandable that under these conditions teachers see little room for time consuming student activities like hands-on experiments.

\item {\bf Insecurity and lack of training:} Even veteran teachers may feel uncomfortable overseeing a class of autonomously experimenting groups – especially, if they have never done so before. The unfamiliar situation requires different classroom management skills. Moreover, when students are truly engaged in an experiment, questions will arise, which the teacher may not be able to answer. While admitting not to know the answer is essential for scientific progress, it is perceived at odds with the teacher's cultural role as authority. For teachers with limited content knowledge, it is therefore easier to align their content closely with textbooks and concentrate on mathematical exercises: According to the official 2011 school census in Alagoas, a state in Brazil's poorer North-East, more mathematicians were teaching physics than actual physicists. Only 20~\% of the physics classes at public schools were given by teachers with (or at the time pursuing) a tertiary degree in physics~\cite{alagoas}. Similar problems were reported in other Brazilian states~\cite{BARRETTO2015}.

\item {\bf Lack of resources and infrastructure:} This is often more an imaginary problem than a real one. Scientific experiments are believed to require sophisticated tools and specialised laboratories. In order to protect such expensive equipment from accidents, students are then only allowed to operate them following detailed step-by-step instructions, instead of following their scientific curiosity. Teachers looking for alternatives will find a variety of online sources proposing experiments with household materials. These experiments, however, are usually designed for entertainment, not for teaching. Placing the teacher in the role of a 'magician', they value impressive visual or acoustic effects above clarity, and may need time consuming preparation. Even if they are considered safe when conducted under adult supervision, this might no longer hold true in a setting where the adult has to keep an eye on 40 to 50 experimenting students at once.
\end{itemize} 

Another reason, why science teachers tend not to include hands-on experiment in their lessons, is the local custom to separate experimental activities from theory lessons. In particular at private schools, laboratory work is often conducted under the supervision of specialised educators and only loosely aligned with the respective theoretical content taught by the physics or chemistry teacher. It might take weeks, before students conducting an experiment will discuss with their teacher, which physical phenomena or chemical reaction had caused the observed effects. Or they might have talked about an effect weeks before getting to see it.

\section{Integrating hands-on research activities in day-to-day teaching}
An initiative like Ciênsação will not be able to remove all these obstacles, but it can help teachers to overcome them. In practice, this means to convince teachers that short experiments – which may take just a couple of minutes or even seconds to conduct – can smoothly transit to productive class discussions, in which students simultaneously advance their fact knowledge, deepen their understanding and foster their scientific skills~\cite{Hattie}. This is based on the assumption that teachers who made positive experiences with this approach, will adapt it to their personal teaching style and make it part of their day-to-day teaching practice.

Teacher training programs are an effective method to convey this idea, but they are limited in their reach by human and financial resources. In order to provide the material to a wider audience, the Ciênsação project has therefore been implemented as an online repository of Open Educational Resources. Hands-on experiments are published at the repository in Portuguese, Spanish and English, so that they are accessible for teachers in all of Latin America. These resources are not only open in the sense that they are freely available under a Creative Commons license, but also in the sense that anyone can contribute by submitting her or his own experiment (which will be peer-reviewed). Professional educators need such editorial freedom to adapt course material to their students' needs. By sharing the content they create in the process, they shape Ciênsação's development and help it adjust dynamically to the community's interests~\cite{Marcos}. 

Published experiments are designed to easily integrate in regular lessons in normal class rooms. Of primary concern is thereby the time and effort required to prepare and conduct an experiment. The less time a teacher has to invest, the more he will be open to try a new methodology. Well-known experiments have therefore been modified to minimise their cost in teaching time, which often goes hand in hand with simplifying them. 

\begin{figure}
	\centering
	\includegraphics[width=0.66\linewidth,clip=true]{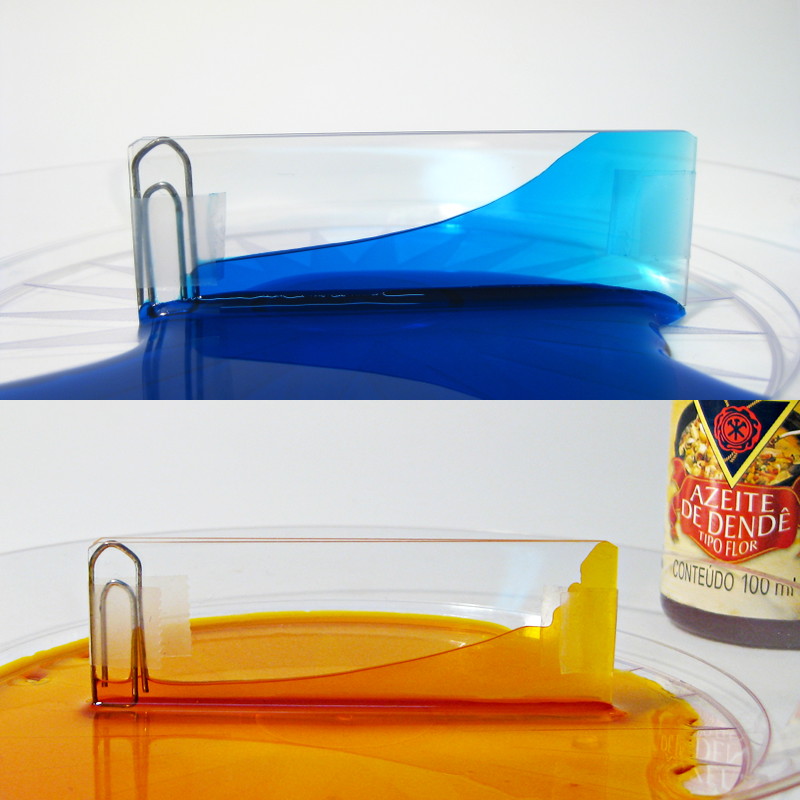}
	\caption[capillary]{Simplifying experiments often means to work with qualitative, rather than quantitative results. In this example, capillary action is studied by comparing its effect on water (blue) and vegetable oil (orange) for different gap width in a wedge made from two cover glasses.
}
	\label{fig:capillary}
\end{figure}

In many cases, this means to work with qualitative results obtained by direct observations, instead of quantitative measurements. Besides reducing the activity's complexity, the concentration on qualitative results also serves the actual purpose of Ciênsação experiments, namely to engage students in an evidence-based discussion about the interpretation of their observations. To embed experiments in a meaningful learning context, each one is built around a specific student research task. The task's questions or challenges provide a clear orientation, what the student is supposed to find out and learn from the experiment. And yet, at times, even apparently simple questions, like ``Are the two ends of a magnet different?'', may lead to sprawling class discussions, challenging teachers to find the fine line between restraining the discussion and allowing students to develop their ideas. Mastering this balance is of great importance for the overall effectiveness and quality of the learning experience. To avoid frustration for both students and teachers, Ciênsação therefore offers 'guiding questions', which break down a student research task in smaller steps, and, when used in a Socratic style, can advance the discussion in the intended direction. Furthermore, these questions can serve teachers to make a smooth transition from an experiment to the respective theoretical background without loosing the motivational momentum of the activity. 

Ciênsação experiments also include a concise explanation of the observed phenomena, which are meant as a starting point for more detailed investigations. The intention is not to replace textbooks, but to reduce preparation time, and to help less experienced teachers in leading the class discussion with confidence.

Materials used in such hands-on experiments need to be selected with great care. No responsible teachers will accept activities which could put their students' health at risk, nor can they be expected to buy expensive equipment, which their school is unable to provide. Consequently, Ciênsação experiments need to rely on safe and easily accessible low-cost materials. Wherever possible, these are household items, ideally objects that students have at hand, like books, pencils or coins. On one hand, simple, generic and versatile materials have the advantage of allowing students to experiment more freely, and to follow their scientific curiosity in unforeseen directions~\cite{photonicsExplorer}. On the other hand, variations in the employed equipment can cause confusion among students, and setups which require much tinkering distract students' attention from the actual phenomena towards the preparations – besides costing teaching time. Before publishing an experiment at the repository, it is therefore tested with a variety of setups to ensure safety and to find the right balance between simplicity, flexibility and material costs.

\begin{figure}
	\centering
	\includegraphics[width=0.5\linewidth,clip=true]{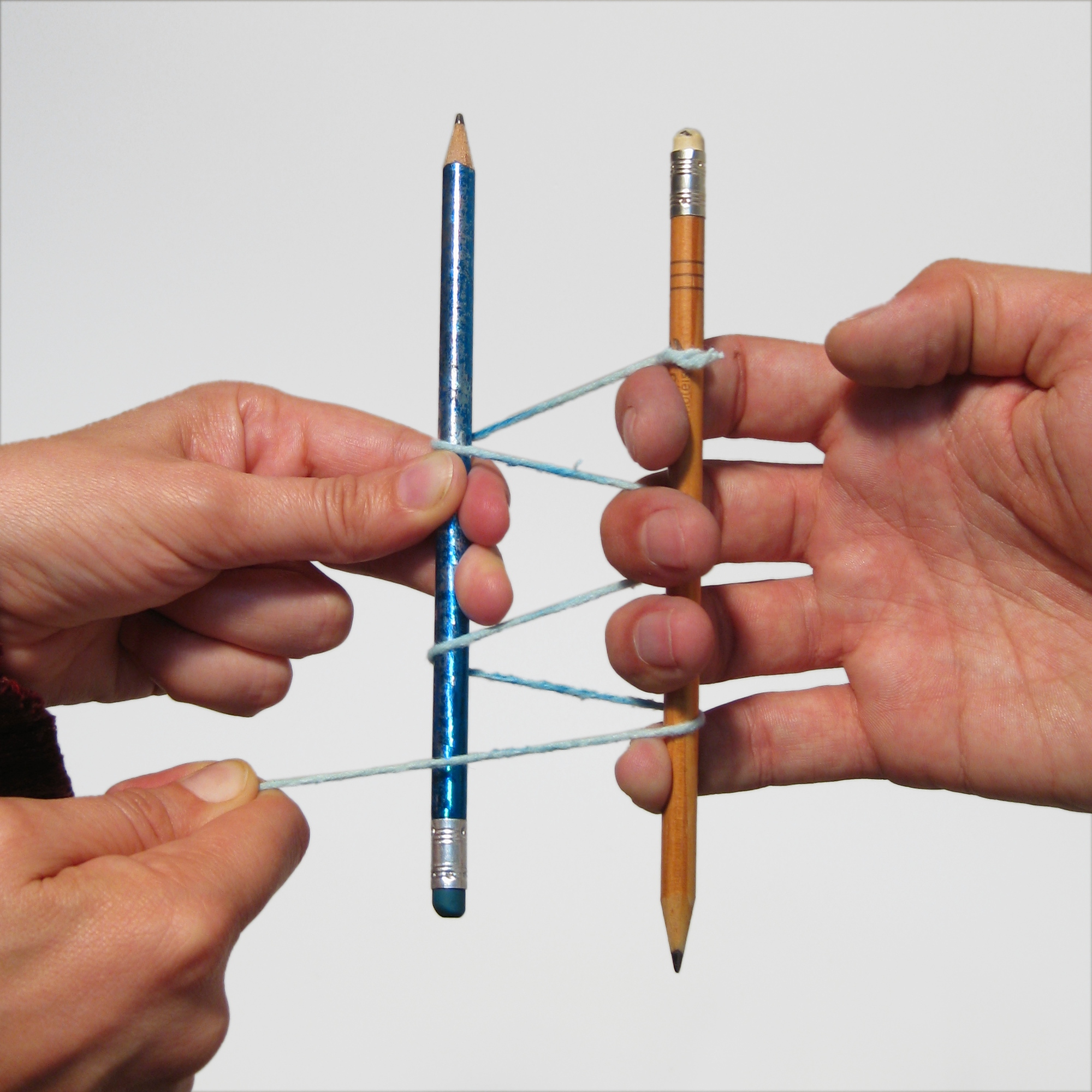}
	\caption[pulley]{To gain a feeling, or intuitive understanding, of mechanical systems like a pulley, students need to experience the involved forces with their own senses. As the example of the pulley shows, this can be achieved with simple means which students usually have at hand anyway.
}
	\label{fig:pulley}
\end{figure}

\section{Gaining a feeling for sciences}
Beyond just making is easy for teachers to integrate them in regular teaching, Ciênsação experiments are specifically designed to foster scientific skills, such as critical thinking, evidence based argumentation and experimental inquiry. Moreover, they are intended to let students gain a feeling for sciences, which gave the project its name: Ciênsação is a neologism joining science (Portuguese: ciência) and feeling (Portuguese: sensação; the English website is named accordingly sciensation.org).

Obviously, responsible individuals should never let subjective feelings interfere when evaluating scientific evidence. However, the common association of science with 'hard' facts often (unintentionally) implies that science is primarily a collection of known facts. This reduces science, which is actually less about facts than about the skill, methodology and joy of finding them, to just a subset of its results. Equally, one could reduce the concept of art to `things found in museums'.

The research tasks, around which Ciênsação experiments are built, invite students to actively \textit{do} science, instead of merely reproducing known results and confirming textbook claims. Giving students a few minutes to pursue such a task autonomously in small groups, allows them not only to experience the excitement of discovery, but also to experience science as a creative activity, as a craft they can master, rather than the privilege of an elite called `scientists'. 

The idea of gaining a feeling for sciences also extends to acquiring an intuitive understanding of physical or chemical systems and their behaviour. It makes a great cognitive difference to either calculate the resonance frequency of a theoretical mass-spring system, or to make a weight oscillate on a spring hold by one's own hand~\cite{expSprings}. Physical laws often make more `sense' to a learner, if they can be related to a personal haptic, visual and acoustic sensation.  

\begin{figure}
	\centering
	\includegraphics[width=0.5\linewidth,clip=true]{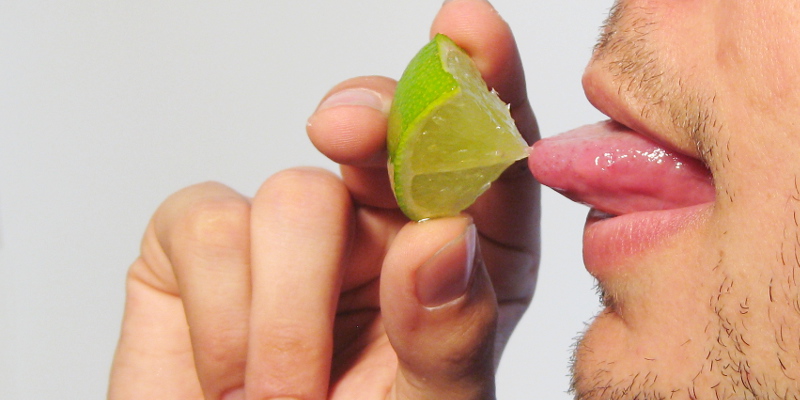}
	\caption[tongue]{Scrutinising textbook claims like the tongue map (referring to different tastes being perceived in distinct regions on the tongue) in experiment can help students to develop an evidence based intellectual autonomy. 
}
	\label{fig:tongue}
\end{figure}

Similarly, students can learn important lessons about the nature of science when discovering contradictions between textbooks and their own experimental results. Errors in some school books – such as claiming that different tastes are perceived in distinct regions on the tongue~\cite{expTongueMap}, or depicting the magnetic north pole on the northern hemisphere~\cite{expNorth} – provide valuable opportunities to teach the importance of searching scientific evidence instead of uncritically relying on authorities. Counter-intuitive observations, like the attraction of equal magnetic poles if a coin is placed between them~\cite{expMagnets}, are often more effective to hone students' research skills than experiments with predictable outcome. Some Ciênsação experiments even discuss the scientific method itself~\cite{expScientificMethod}. To show that a majority opinion does not equal scientific evidence, one experiment asks students to observe the movement of laser speckles on a wall while moving their heads~\cite{expSpeckle}. Whether the speckle pattern moves in the same or the opposite direction as their head, depends on each observer's individual eyesight (observers with hyperopia will see the speckle move in the direction of the head movement, while myopic observers will see them moving in the opposite direction). Nonetheless, the teacher – temporarily deceiving her students – pressures the class to decide on one direction and contests diverting opinions, before revealing the true nature of the experiment. If some students started to doubt their own observation or even gave in to peer pressure, the whole class gets to know a critical pitfall of the scientific endeavour.

\section{Impact}
Whether the project Ciênsação can meet the ambitious expectations set in it, is currently tested in teacher training courses at the Federal University of Paraná (Brazil) and the Federal University of Fronteira Sul (Santa Catarina, Brazil). A recent workshop, held at the Federal University of Rio Grande do Sul, brought together educational researchers, science teachers and students to discuss how Ciênsação can be extended to reach more teachers, and allow them to interact more efficiently with the platform and each other. 
 
Meanwhile, experiments in the online repository are read by teachers from all over the world, and comments of appreciation were received from countries as diverse as Ghana and Germany. An offline version of the experiments is distributed on memory sticks to teachers on Fiji and Solomon Islands, who have limited access to the internet. This encouraging feedback and a growing number of online visitors indicate that the first steps of the Ciênsação project indeed contribute to a culture of short hands-on experiments and skill-fostering science education – well beyond the intended audience of teachers in Latin America. 

\section{References}
\bibliography{CSC4IOP}
\bibliographystyle{IEEEtran}